\documentclass[aps,prd,superscriptaddress,nofootinbib,11pt]{revtex4}
\usepackage[english]{babel}
\usepackage[utf8]{inputenc}
\usepackage{graphicx}   
\usepackage{slashed}
\usepackage{epstopdf}
\usepackage{verbatim}   
\usepackage{color}      
\usepackage{subfigure}  
\usepackage{multirow}
\usepackage{hyperref}   
\usepackage{float}
\usepackage{epsfig,rotating}
\usepackage{amsmath,amssymb}
\usepackage{dsfont}
\usepackage{slashed}
\restylefloat{table}
\numberwithin{equation}{section}

\newcommand{\vx}{\vec{x}}
\newcommand{\vp}{\vec{p}}
\newcommand{\vq}{\vec{q}}
\newcommand{\vk}{\vec{k}}

\newcommand{\op}{\mathcal{O}_\chi}
\newcommand{\vy}{\vec{y}}

\newcommand{\be}{\begin{equation}}
\newcommand{\ee}{\end{equation}}
\newcommand{\bea}{\begin{eqnarray}}
\newcommand{\eea}{\end{eqnarray}}

\newcommand{\ket}[1]{|#1\rangle}
\newcommand{\bra}[1]{\langle#1|}

\begin{document}
\title{ Non-equilibrium dynamics of  Axion-like particles:\\ the quantum master equation.}

\author{Shuyang Cao}
\email{shuyang.cao@pitt.edu} \affiliation{Department of Physics and
Astronomy, University of Pittsburgh, Pittsburgh, PA 15260}
\author{Daniel Boyanovsky}
\email{boyan@pitt.edu} \affiliation{Department of Physics and
Astronomy, University of Pittsburgh, Pittsburgh, PA 15260}

 \date{\today}

\begin{abstract}

We study the non-equilibrium dynamics of Axion-like particles (ALP) coupled to Standard Model degrees of freedom in thermal equilibrium. The  Quantum Master Equation (QME)  for the (ALP) reduced density matrix is derived to leading order in the coupling of the (ALP) to the thermal bath, but to \emph{all} orders of the bath couplings to degrees of freedom within or beyond the Standard Model other than the (ALP).  The (QME) describes the damped oscillation dynamics of an initial misaligned (ALP) condensate, thermalization with the bath, decoherence and entropy production within a unifying framework. The (ALP) energy density $\mathcal{E}(t)$ features two components: a ``cold'' component from the misaligned condensate and a ``hot''   component from thermalization with the bath,  with $\mathcal{E}(t)= \mathcal{E}_{c}\,e^{-\gamma(T)\,t}+\mathcal{E}_{h}(1-e^{-\gamma(T)\,t})$ thus providing a ``mixed dark matter'' scenario.     Relaxation of the (ALP) condensate, thermalization, decoherence and entropy production  occur on similar time scales. An explicit example with  (ALP)-photon coupling,  valid post recombination yields a relaxation rate $\gamma(T)$ with a substantial enhancement from thermal emission and absorption.  A misaligned condensate is decaying at least since recombination and on the same time scale thermalizing with the cosmic microwave background (CMB). Possible consequences for   birefringence of the (CMB) and (ALP) contribution to the effective number of ultrarelativistic species and galaxy formation are discussed.
\end{abstract}


\maketitle

\section{Introduction}

The axion,  introduced in Quantum Chromodynamics (QCD) as a solution of the strong CP problem\cite{PQ,weinaxion,wil} may be produced non-thermally in the Early Universe, for example by a misalignment mechanism and  is recognized as a potentially viable  cold dark matter candidate\cite{pres,abbott,dine}. Extensions beyond the standard model can accommodate pseudoscalar particles with properties similar to the QCD axion, namely axion-like-particles (ALP) which can also be suitable dark matter candidates\cite{banks,ringwald,marsh,sikivie1,sikivie2}, in particular as candidates for ultra light dark matter\cite{fuzzy,uldm}. Constraints on the mass and couplings of ultra light (ALP)\cite{marsh,sikivie1,sikivie2,banik} are being established  by various experiments\cite{cast,admx,graham}. There are two important features that characterize (ALP), i) a misalignment mechanism results in damped coherent oscillations of the expectation value of the (ALP) field which gives rise to the contribution to the energy density as a cold dark matter component\cite{pres,abbott,dine,marsh,sikivie1,sikivie2,turner}, ii) its pseudoscalar nature leads to an interaction between the (ALP) and photons  or  gluons via pseudoscalar composite operators of gauge fields, such as $\vec{E}\cdot\vec{B}$ in the case of the (ALP)-photon interaction and $G^{\mu\nu;b}\widetilde{G}_{\mu\nu;b}$ in the case of gluons, which allows an (ALP) to decay into two photons or gluons. The effect of this decay process in the evolution of (ALP) condensates has been studied in refs.\cite{khlopov,sigl,arza,dashin} including stimulated decay in a photon background. The damping of an (ALP) condensate via a ``friction'' term in its equation of motion has been studied in refs.\cite{mottola,friction,friction1}, and thermalization of (ALP) has been studied in refs.\cite{buch,masso}, these references focused on either damping via friction or thermalization as unrelated independent processes.  A recent study\cite{shuyang} has recognized the common origin of these two seemingly different processes by obtaining the non-equilibrium effective action that determines the time evolution of the reduced (ALP) density matrix. This study showed that damping of a misaligned (ALP) condensate and thermalization are two complementary aspects and are linked by the fluctuation dissipation relation, a fundamental   and ubiquitous property of a bath in thermal equilibrium. This reference also established that both processes contribute to the (ALP) energy density, an important aspect if the (ALP) are suitable dark matter candidates.

Decay and thermalization of an (ALP) condensate post recombination may have profound impact on birefringence of the cosmic microwave background (CMB) if its origin is the electromagnetic coupling of a pseudoscalar (ALP)\cite{sikibiref,komatsu,mina,biref}.

   In ref.\cite{shuyang}, the non-equilibrium dynamics of (ALP) was studied to leading
order in the coupling of the (ALP) to other degrees of freedom treated as a bath  in thermal equilibrium by implementing the in-in Schwinger-Keldysh formulation of non-equilibrium quantum field theory to obtain the effective action. The   equations of motion for the (ALP) obtained from the effective action are retarded and akin to
a Langevin equation with a friction term determined by the retarded self-energy and a noise term related to the self energy via the fluctuation -dissipation relation. This relation is a consequence of the bath degrees of freedom being in thermal equilibrium. An important result of the Langevin nature of the effective equations of motion is a direct relationship between the damping of an (ALP) coherent condensate and thermalization of its fluctuations. This result was found to be general to leading order in the (ALP) coupling to the bath degrees of freedom but to all orders in the couplings of these ``environmental'' fields to any other field within or beyond the standard model other than the (ALP) and is a  corollary of the fluctuation dissipation relation.  An analysis of the   coupling of (ALP) to the (CMB) post recombination  in this article also revealed   a substantial enhancement of the damping and thermalization rates if the (ALP) is an ultra-light dark matter candidate as well as  unexpected possible phase transitions and exotic new phases.

\vspace{1mm}

\textbf{ Motivation and Objectives:}

The results in ref.\cite{shuyang} and their possible cosmological consequences,  motivate us to seek a complementary formulation of the non-equilibrium dynamics of (ALP) coupled to ``environmental'' degrees of freedom in equilibrium that does not rely on the in-in Schwinger-Keldysh approach to the effective action, thereby offering an alternative and independent assessment of the non-equilibrium dynamics of (ALP) coupled to a thermal environment.

   In this article we adapt methods of quantum optics and quantum information to study the  non-equilibrium dynamics of (ALP) fields implementing a quantum master equation approach ubiquitous in the treatment of quantum open systems\cite{breuer,zoeller,lin,gori,pearle}. The quantum master equation describes the time evolution of the (ALP) reduced  density matrix,   it has been implemented in particle physics\cite{weinberg1,banks2,openburra,openaka,openyao,openmiura,openbram} and cosmology\cite{boyopen,hollow,shandera,berera,vennin,bartolo} and has proven to be a powerful and reliable method to study non-equilibrium dynamics.

    The main objectives of this article are: \textbf{i)} to scrutinize the results obtained in ref.\cite{shuyang} with an alternative and independent method, \textbf{ii)} to inquire on complementary aspects of the time evolution of the reduced density matrix, in particular the evolution of coherences, which yield supplementary information on thermalization and decoherence,  and \textbf{iii)} to compare the time scales of decoherence to those of damping of the misaligned condensate and thermalization.

    In this study we are not concerned with bounds on couplings and or masses of the putative (ALP) but focus on fundamental aspects of the non-equilibrium evolution of  its density matrix including   misaligned initial states. Furthermore, while our ultimate objective is to study the non-equilibrium dynamics in an expanding cosmology,   we initiate this program as a prelude by focusing on Minkowski space time.

    \vspace{1mm}

    \textbf{Brief summary of results:} We consider an (ALP) field in interaction with Standard Model degrees of freedom which are considered to be in thermal equilibrium. In section (\ref{sec:master}), we obtain the (QME)  for the reduced density matrix of the (ALP) up to second order in the coupling of the (ALP) to these degrees of freedom, but to all orders in the coupling of the bath degrees of freedom to fields within or beyond the Standard Model different from the (ALP)  under well defined approximations. The  resulting (QME) is of the Lindblad form\cite{breuer,zoeller,lin,gori,pearle}, it  is obtained up to second order in the (ALP) coupling to Standard Model degrees of freedom and  to all orders in the couplings of these degrees of freedom to any other field within or beyond the Standard Model except for the (ALP), and includes misaligned initial conditions for the (ALP) field. The (QME) describes the damping of the misaligned condensate, thermalization with the bath and decoherence with a concomitant entropy production. The (ALP) energy density describes a ``mixed'' dark matter scenario with a cold component $\mathcal{E}_c$ from the misaligned coherent condensate, and a ``hot'' component $\mathcal{E}_h$ from thermalization with the bath, with the total energy density interpolating between the cold and hot components as $\mathcal{E}(t)\simeq  \mathcal{E}_c\,e^{-\gamma(T)t}+ \mathcal{E}_h\,(1-e^{-\gamma(T)t})$, where the relaxation rate $\gamma(T)$ also describes the decoherence rate. We study in detail (ALP) coupling to the (CMB) post recombination, if the (ALP) is an ultralight dark matter candidate there is a substantial enhancement of the relaxation rate, its long wavelength limit is given by $\gamma(T) = g^2 m^2_a \, T/16\pi$. The results suggest that if $\gamma(T) < 1/H_0$ the misaligned condensate has been decaying at least since recombination and thermalizing with the (CMB) on a similar time scale. Therefore, if cosmic birefringence is a consequence of the (CMB) coupling to a pseudoscalar (ALP), the rotation angle since the surface of last scattering should feature a thermal spectrum of fluctuations.

\vspace{1mm}

 \section{The quantum Master equation:}\label{sec:master}

  We study  the time evolution of the reduced density matrix of  an axion-like field $ a(x)$ coupled to   generic fields $\chi(x)$ to which we refer as ``environmental'' fields via a pseudoscalar operator $\op(x)$, with the Lagrangian density
  \be \mathcal{L}[a,\chi] = \frac{1}{2}\,\partial_\mu a(x) \partial^\mu a(x) - \frac{1}{2}\,m^2_{a}\,a^2(x) - g a(x)\,\op(x) + \mathcal{L}_{\chi} \label{lag}\ee where $\mathcal{L}_{\chi}$ is the Lagrangian density describing the ``environmental''  fields $\chi$, these fields could be the electromagnetic field, fermion or gluon fields and themselves be coupled to other degrees of freedom within or beyond the Standard Model.

   The Lagrangian density (\ref{lag}) describes several relevant couplings of (ALP), with possible operators $\op(\vx)$ being
  $\op(\vx) =   \vec{E}(x)\cdot \vec{B}(x)~;~     G^{\mu \nu,b}(x)\widetilde{G}_{\mu \nu,b}(x)  ~;~   \overline{\Psi}(x)\gamma^5 \Psi(x) \cdots $ where $\vec{E},\vec{B}$ are the electromagnetic fields, $G^{\mu \nu,b};\widetilde{G}^{\mu \nu,b}$ are the gluon field strength tensor and its dual respectively, and $\Psi(x)$ a fermionic field. These degrees of freedom   are assumed to be in thermal equilibrium. We  will first treat these fields generically denoting them as $\chi$ fields, and after obtaining the general form of the quantum master equation up to $\mathcal{O}(g^2)$,   we will   focus on the relevant case with $\op(\vx) =  \vec{E}(x)\cdot \vec{B}(x)$ since the interaction of (ALP) fields with the (CMB) could have potentially observable consequences, such as  birefringence\cite{sikibiref,komatsu,mina,biref},  a rotation of the polarization plane which, in contrast to Faraday rotation,  is independent of the frequency with tantalizing detection possibilities\cite{komatsu,mina,biref}.

  The  interaction of (ALP) with photons and gluons  via couplings of the form $g a(x)\vec{E}(x)\cdot \vec{B}(x);g_s a(x)G^{\mu \nu,b}(x)\widetilde{G}_{\mu \nu,b}(x)   $  are not renormalizable because the respective couplings $g,g_s$ feature dimensions $1/(energy)$, an aspect that has important consequences\cite{shuyang} discussed below, that at the fundamental level, indicate that the Lagrangian density (\ref{lag}) describes an effective field theory valid below some cutoff scale.

Upon evolving the total initial density matrix in time, the degrees of freedom $\chi$  with the generic operator $\op$ are traced over to obtain a reduced density matrix for $a(x)$ which obeys a quantum master equation. We obtain this equation in the general case valid to order $g^2$ in the (ALP) coupling to the bath, and to all orders in the couplings of the bath degrees of freedom to any other degree of freedom within or beyond the standard model except for the (ALP) under a set of approximations that are spelled out in detail. Whereas our ultimate objective is to pursue this approach in an expanding cosmology, here we begin this program by first carrying it out in Minkowski space time.

 The quantum master equation in a Lindblad form\cite{lin,gori,pearle,weinberg1}   has   recently received attention in applications to high energy physics\cite{banks2,openburra,openaka,openyao,openmiura,openbram} and cosmology\cite{boyopen,hollow,shandera,berera,vennin}. This formulation begins with the time evolution of an initial density matrix that describes the total system of fields $a,\chi$, which is given by
  \be \hat{\rho}(t) = e^{-iHt}\hat{\rho}(0)e^{iHt}\,, \label{rhooft} \ee with $H$ the total Hamiltonian
  \be H = H_{0a}+ H_{\chi}+ H_I \equiv H_0+H_I \,, \label{totham}\ee where $H_{0a}$ is the free field Hamiltonian for the (ALP), $H_{\chi}$ is the Hamiltonian of the $\chi$ degrees of freedom  including their couplings to degrees of freedom within or beyond the Standard Model except the (ALP), and $H_I= g \,
  \int d^3 x  a(x)\,\mathcal{O}_{\chi}(x)$ is the coupling between the (ALP) and the bath degrees of freedom obtained from the Lagrangian density (\ref{lag}).

 We consider an initial factorized density matrix
  \be \rho(0) = \rho_a(0)\otimes \rho_{\chi}(0)\,, \label{inirhofac}\ee  where the $\chi$ fields are in thermal equilibrium at temperature $T=1/\beta$, namely
  \be \rho_{\chi}(0) = \frac{e^{-\beta H_{\chi}}}{\mathrm{Tr}e^{-\beta H_{\chi}}}\,, \label{rhother} \ee and
  for the (ALP) field we chose an initial density matrix describing a ``misaligned'' initial condition with a non-vanishing expectation value of the (ALP) field. This  is implemented in terms of coherent states of free fields as follows. Quantizing  the free (ALP) field at the initial time $t=0$  in a finite volume $V$ as
  \be a(\vx,t=0) = \frac{1}{\sqrt{V}}\sum_{\vk} \frac{1}{\sqrt{2\omega_k}}\Big[b_{\vk}\,  e^{i \vk\cdot \vx}+b^\dagger_{\vk}\,  e^{-i \vk\cdot \vx}  \Big]~~;~~ [b_{\vk},b^\dagger_{\vk'}] = \delta_{\vk,\vk'}  \,,\label{aquant}\ee and the vacuum state defined as
  \be b_{\vk}\ket{0} =0 \,.\label{vacuum} \ee  A coherent state is given by
\be \ket{\Delta} = \Pi_{\vk}\, e^{- \frac{1}{2}|\Delta_{\vk}|^2}\,e^{-\Delta_{\vk} b^\dagger_{\vk}}\,\ket{0}\,, \label{cs} \ee it is an eigenstate of the annihilation operator,
\be b_{\vk} \ket{\Delta} = \Delta_{\vk} \ket{\Delta} \,,  \label{eigenb}\ee  and describes a Poisson distribution of quanta of the free (ALP) field.  The expectation values of the (ALP) field and its canonical momentum in this coherent state are
\bea \bra{\Delta} a(\vx,0) \ket{\Delta}  & = & \overline{a}(\vx,0) = \frac{1}{\sqrt{V}}\sum_{\vk}\frac{1}{\sqrt{2\omega_k}} \Big[\Delta_{\vk}   +\Delta^{*}_{-\vk}    \Big]\, \,e^{i \vk\cdot \vx}\,,\label{expealp}\\
 \bra{\Delta} \pi(\vx,0) \ket{\Delta}  & = & \overline{\pi}(\vx,0) = \frac{-i}{\sqrt{V}}\sum_{\vk} \sqrt{\frac{\omega_k}{2 }}\Big[\Delta_{\vk}   -\Delta^{*}_{-\vk}   \Big]\,  e^{i \vk\cdot \vx}\,, \label{expepialp}
\eea
Hence we choose the initial density matrix for the (ALP) field to describe this ``misaligned'' initial state, namely
\be \rho_{a}(0) =  \ket{\Delta}\bra{\Delta}\,,  \label{rhozeroalp}\ee yielding
\be \mathrm{Tr} b_{\vk} \rho_{a}(0) = \Delta_{\vk}~~;~~ \mathrm{Tr} b^\dagger_{\vk}b_{\vk} \rho_{a}(0)= N_q(0) = |\Delta_{\vq}|^2  ~~;~~ \mathrm{Tr} b_{\vk}b_{-\vk} \rho_{a}(0) = \Delta_{\vk}\Delta_{-\vk}~~;~~\mathrm{etc.} \,. \label{trazasbb}\ee

We refer to the off-diagonal (ALP) density matrix elements in the occupation number basis (eigenstates of  $b^\dagger_{\vk}b_{\vk}$), for example $\mathrm{Tr} b_{\vk}b_{-\vk} \rho_{a}(0)=\Delta_{\vk}\Delta_{-\vk}$  as \emph{coherences}\cite{breuer,zoeller}. A hallmark of a thermal density matrix is that these coherences vanish and the density matrix is diagonal in the occupation number basis. This observation will become important as a diagnosis of thermalization  and its link to decoherence studied below.

Translational invariance entails that
\be \Delta_{\vk} = \sqrt{V}\, \widetilde{\Delta}  \,\delta_{\vk,\vec{0}}\,,  \label{deltazero}
\ee  therefore
\be N_q(0) = V |\widetilde{\Delta}|^2 \delta_{\vq,\vec{0}}\,, \label{Nzero} \ee and
\be \frac{1}{2m_a}\Big[ {\overline{\pi}^2}  +  m^2_a\,\overline{a}^2\Big] = |\widetilde{\Delta}|^2\,. \label{zeromodi} \ee

  In the quantum master equation approach\cite{breuer,zoeller} the time evolution of the density matrix is considered in the interaction picture. With  the full density matrix $\hat{\rho}(t)$  given by eqn. (\ref{rhooft}) the density matrix in the interaction picture is given by
\be \hat{\rho}_I(t)= e^{iH_0 t} \hat{\rho}(t) e^{-iH_0t}\,,\label{rhoIP}\ee whose time evolution obeys
\be \dot{\hat{\rho}}_I(t) = -i \big[H_I(t),\hat{\rho}_I(t)\big]\,, \label{rhodotip}\ee where $H_I(t)$ is the interaction Hamiltonian   in the interaction picture, $H_I(t)= e^{iH_0t}H_I e^{-iH_0t}$.   The formal solution of eqn. (\ref{rhodotip})  is given by
\be \hat{\rho}_I(t)= \hat{\rho}_I(0) -i \int^t_0 dt' \big[H_I(t'),\hat{\rho}_I(t')\big] \,. \label{solurhoip}\ee This solution is inserted back into (\ref{rhodotip}) leading to the iterative equation
\be \dot{\hat{\rho}}_I(t) = -i \big[H_I(t),\hat{\rho}_I(0)\big] -\int^t_0   \,\big[H_I(t),\big[H_I(t'),\hat{\rho}_I(t')\big]\big]\,dt' \,.\label{rhodotiter}\ee
 This (QME)  cannot be solved exactly, and several approximations are usually invoked, based on the following \emph{assumptions}\cite{breuer,zoeller,lin,gori}:

\begin{itemize}

\item \textbf{\underline{Factorization}}: the total density matrix factorizes into a direct product of the density matrix for the $a$ field, $\hat{\rho}_{Ia}(t)$ and that of the bath of $\chi$ fields, $\hat{\rho}_\chi$, namely,
\be \hat{\rho}_I(t) = \hat{\rho}_{I a}(t)\otimes \hat{\rho}_\chi(0)\,, \label{factorization}\ee  where
\be \hat{\rho}_\chi(0)= \frac{e^{-\beta H_{ \chi}}}{\mathrm{Tr}e^{-\beta H_{ \chi}}}\,,\label{rhozerochi}\ee this  \emph{assumption} which implies that the   bath degrees of freedom   remain in thermal equilibrium, relies on that the bath is a reservoir with a large number of degrees of freedom and is not modified by its coupling to the system, hence the density matrix of the bath does not depend on time. This assumption also relies on weak coupling: if the initial density matrix is factorized, correlations between the system and the reservoir will build as a consequence of  the interaction, therefore such correlations will be small for very weak coupling and may only contribute in higher orders. Factorization and its possible caveats are discussed further in section (\ref{sec:discussion}).

The \emph{reduced density matrix} for the (ALP) field $a$ is obtained by taking the trace of the full density matrix over the bath degrees of freedom, which by assumption remains in thermal equilibrium, therefore
\be  \hat{\rho}_{Ia}(t)  = \mathrm{Tr}_{\chi} \hat{\rho}_I(t) \,. \label{redmtxfi}\ee

 Upon taking the trace over the $\chi$ degrees of freedom  the first term on the right hand side of eqn. (\ref{rhodotiter}) vanishes under the assumption that the thermal density matrix of the environmental fields is even under parity, hence $\mathrm{Tr}\mathcal{O}_{\chi}\,\hat{\rho}_{\chi}(0) =0$, and we find the evolution equation for the  reduced density matrix for the (ALP) field $a$ in the interaction picture,
\bea \dot{\hat{\rho}}_{Ia}(t)   & =  & -g^2\int^t_0 dt' \int d^3x \int d^3 x'\Bigg\{ a_I(x) \, a_I(x')\,\hat{\rho}_{Ia}(t') \,\,G^>(x-x') + \hat{\rho}_{Ia}(t') \,a_I(x')\, a_I(x)\,G^<(x-x') \nonumber \\
& - & a_I(x) \,\hat{\rho}_{Ia}(t') \, a_I(x')\, G^<(x-x') - a_I(x') \, \hat{\rho}_{Ia}(t')\, a_I(x)\,G^>(x-x') \Bigg\} \label{Linblad}\eea
where   we use the shorthand convention $x \equiv (\vec{x},t)~;~x' \equiv (\vec{x}',t')$, and introduced the bath correlation functions
\bea  G^>(x-x') & = &  \mathrm{Tr}_{\chi} \hat{\rho}_\chi(0) \mathcal{O}_{\chi}(x)  \mathcal{O}_{\chi}(x') \label{ggreat} \\
G^<(x-x') & = &  \mathrm{Tr}_{\chi} \hat{\rho}_\chi(0) \mathcal{O}_{\chi}(x')\mathcal{O}_{\chi}(x)  \,.  \label{gless} \eea
The (ALP) field in the interaction picture $a_I(\vx,t)$ features free field time evolution, namely
\be a_I(\vx,t)= \frac{1}{\sqrt{V}}\sum_{\vk} \frac{1}{\sqrt{2\omega_k}}\Big[b_{\vk}\, e^{-i\omega_k t} \,e^{i \vk\cdot \vx}+b^\dagger_{\vk}\,e^{i\omega_k t}\,  e^{-i \vk\cdot \vx}  \Big]\,,\label{alpip}\ee where the operators $b_{\vk},b^\dagger_{\vk}$ do not depend on time, and $\omega_k=\sqrt{k^2+m^2_a}$.

\item   \underline{\textbf{Markov approximation}} the second approximation entails replacing $\rho_{Ia}(t') \rightarrow \rho_{Ia}(t)$ in the time integral. This is usually referred to as a Markov approximation and   is  justified in weak coupling, as can be seen by   considering the first term in (\ref{Linblad}) as an example. It can be written as
\be -g^2 a(\vx,t) \int^t_0 \frac{d\mathcal{K}(t')}{dt'} \,\hat{\rho}_{Ia}(t') \,dt' ~~;~~  \mathcal{K}(t') \equiv \int^{t'}_0 a({\vx}',t'') \, G^>(\vx-{\vx}',t-t'')dt'' \label{incha}\ee which upon integration by parts yields
\be -g^2 a(\vx,t)  \mathcal{K}(t) \hat{\rho}_{Ia}(t) + g^2  a(\vx,t) \int^t_0  \mathcal{K}(t') \,\frac{d\hat{\rho}_{Ia}(t')}{dt'} dt' \label{incha2}\ee in the second term $d\hat{\rho}_{I\Phi}(t')/dt' \propto g^2$ so this   term yields a contribution that is formally of order $g^4$ and can be neglected to second order. The same analysis is applied to all the other terms in (\ref{Linblad}) with the conclusion that in weak coupling and  to leading order $(g^2)$  the Markovian approximation  $\hat{\rho}_{Ia}(t') \rightarrow \hat{\rho}_{Ia}(t)$ is justified.

Therefore in the Markov approximation the quantum master equation becomes
\bea \dot{\hat{\rho}}_{Ia}(t)   & =  & -g^2\int^t_0 dt' \int d^3x \int d^3 x'\Bigg\{ a_I(x) \, a_I(x')\,\hat{\rho}_{Ia}(t) \,\,G^>(x-x') + \hat{\rho}_{Ia}(t) \,a_I(x')\, a_I(x)\,G^<(x-x') \nonumber \\
& - & a_I(x) \,\hat{\rho}_{Ia}(t) \, a_I(x')\, G^<(x-x') - a_I(x') \, \hat{\rho}_{Ia}(t)\, a_I(x)\,G^>(x-x') \Bigg\} \,. \label{markovlim}\eea

The correlation functions $G^{>}(x-x'),G^<(x-x')$ are obtained in appendix (\ref{app:corre}) in terms of non-perturbative  Lehmann  representations to all orders in the coupling of the environmental fields $\chi$ to any other field in thermal equilibrium except  for the (ALP).  They are given by
\bea  G^>(x-x') & = & \int \frac{d^3q}{(2\pi)^3} \int \frac{dq_0}{2\pi} ~\varrho^>(q_0,\vq)\,e^{-iq_0(t-t')}~e^{i \vq\cdot(\vx-\vx')} \label{ggreatspec}\\
G^<(x-x') & = & \int \frac{d^3q}{(2\pi)^3} \int \frac{dq_0}{2\pi} ~\varrho^<(q_0,\vq)\,e^{-iq_0(t-t')}~e^{i \vq\cdot(\vx-\vx')} \,,\label{glessspec}\eea where the spectral densities obey the relation
\be \varrho^>(-q_0,\vq)= \varrho^<(q_0,\vq)\,,\label{rela3}  \ee and fulfill the Kubo-Martin-Schwinger condition\cite{kms}
\be \varrho^<(q_0,\vq) = e^{-\beta\,q_0} \,\varrho^>(q_0,\vq)\,, \label{kms} \ee
which  is  a consequence of   the fields $\chi$ being  in thermal equilibrium. Introducing the spectral density
\be \varrho(q_0, \vq) = \varrho^>(q_0,\vq)-\varrho^<(q_0,\vq)\,,  \label{specdens}\ee the Kubo-Martin-Schwinger condition (\ref{kms}) leads to the following relations
\bea \varrho^>(q_0,\vq) & = &  [1+n(q_0)]\,\varrho(q_0,\vq) \label{rhogreat2} \\\varrho^<(q_0,\vq) & = &   n(q_0)\,\varrho(q_0,\vq) \label{rholess2} \eea where $n(q_0) = [e^{\beta\,q_0}-1]^{-1}$ is the Bose-Einstein distribution function at temperature $T=1/\beta$.  The above relations are proven in appendix (\ref{app:corre}), they are general,   non-perturbative and rely only on that the reservoir is in thermal equilibrium.

\item \underline{\textbf{rotating wave approximation:}}  in writing the products $a_I(\vx,t)~a_I(\vx',t')$ of interaction picture field operators (\ref{alpip}) in (\ref{Linblad}) there are two types of terms  with very different time evolution. Terms of the form
\be b^\dagger_{\vq}~b_{\vq}~e^{  i{\omega_q(t-t')}}\,, \label{ada} \ee   are ``slow'', and terms of the form
\be b^\dagger_{\vq}~b^\dagger_{-\vq}~ e^{2i\omega_q t}~e^{i{\omega_q(t-t')}}~~;~~ b_{\vq}\,b_{-\vq}~ e^{-2i\omega_q t}\,e^{-i{\omega_q(t-t')}}\,, \label{adad}\ee are fast, the extra rapidly varying phases $e^{ \pm 2i\omega_q t}$ lead to rapid dephasing on time scales $\simeq 1/\omega_q$ and do not yield resonant (nearly energy conserving) contributions. Neglecting these terms is tantamount to neglecting non-resonant terms that average out over the longer time scales of relaxation $\gg 1/\omega_q$.   These terms only give perturbatively small transient contributions and are discussed in section (\ref{sec:discussion}). Keeping only the slow terms which dominate the long time dynamics  for $t \gg 1/\omega_q$ and neglecting the fast oscillatory terms defines the ``rotating wave approximation'' ubiquitous in quantum optics\cite{breuer,zoeller}.

\end{itemize}

We will adopt these approximations and comment in section (\ref{sec:discussion}) on the corrections associated with keeping the fast   terms as well as caveats in the factorization approximation and limitations of the (QME).

Implementing the  {Markov}  approximation $\hat{\rho}_{Ia}(t')\rightarrow \hat{\rho}_{Ia}(t)$, and the    {rotating wave } approximation (keeping only terms of the form $b^\dagger~b, b~b^\dagger$) using the spectral representation of the correlators (\ref{ggreatspec},\ref{glessspec})    and carrying out the spatial and temporal integrals we obtain the  \emph{Lindblad} form\cite{breuer,zoeller,lin,gori,pearle,weinberg1} of the quantum master equation,
\bea \dot{\hat{\rho}}_{Ia}(t)  & = &  \sum_{\vq} \Bigg\{   -i\,\Delta_q(t)~\Big[b^\dagger_{\vq}\,b_{\vq}, \hat{\rho}_{Ia}(t) \Big] \nonumber \\
& - & \frac{\Gamma^>_q(t)}{2} \Big[b^\dagger_{\vq}\,b_{\vq}~ \hat{\rho}_{Ia}(t) + \hat{\rho}_{Ia}(t)~b^\dagger_{\vq}\,b_{\vq} - 2 b_{\vq}~\hat{\rho}_{Ia}(t)~b^\dagger_{\vq} \Big]\nonumber \\
& - & \frac{\Gamma^<_q(t)}{2} \Big[b_{\vq}\,b^\dagger_{\vq}~ \hat{\rho}_{Ia}(t) + \hat{\rho}_{Ia}(t)~b_{\vq}\,b^\dagger_{\vq} - 2 b^\dagger_{\vq}~\hat{\rho}_{Ia}(t)~b_{\vq} \Big] \Bigg\} \,,
\label{Linfin} \eea  where
\be \Delta_q(t) = \frac{g^2}{2\omega_q} \,\int \frac{dq_0}{2\pi} \,\varrho(q_0,q)\,\frac{\Big[1-\cos[(\omega_q-q_0)t] \Big]}{(\omega_q-q_0)}\,, \label{Roftim}\ee
\be \Gamma^>_q(t) = \frac{g^2}{ \omega_q}  \,\int \frac{dq_0}{2\pi} \, \varrho(q_0,q)\,\big[1+n(q_0)\big]\frac{ \sin[(\omega_q-q_0)t] }{ (\omega_q-q_0)} \,, \label{gamgre}\ee
\be \Gamma^<_q(t) = \frac{g^2}{\omega_q}  \,\int \frac{dq_0}{2\pi} \, \varrho(q_0,q)\, n(q_0) \frac{ \sin[(\omega_q-q_0)t] }{(\omega_q-q_0)} \,, \label{gamles}\ee
and we introduce
\be \Gamma_q(t) = \Gamma^>_q(t)-\Gamma^<_q(t)= \frac{g^2}{\omega_q} \,\int \frac{dq_0}{2\pi} \,\varrho(q_0,q)\,\frac{ \sin[(\omega_q-q_0)t] }{(\omega_q-q_0)}\,. \label{gamadif}\ee

The second and third lines in (\ref{Linfin}) are called the \emph{dissipator}\cite{breuer}, these are non-Hamiltonian, purely dissipative terms, however it follows from the (QME) (\ref{Linfin}) that the trace of the reduced density matrix is conserved. It is argued in refs. \cite{lin,gori,pearle,weinberg1}   that the equation (\ref{Linfin}) is the most general linear evolution equation that preserves unit trace and  Hermiticity of the density matrix.

 Expectation values of  (ALP) operators in the interaction picture are obtained by taking the trace of such operators with the reduced density matrix, for example

\be \langle a_I(\vx,t) \rangle = \mathrm{Tr}\, a_I(\vx,t) \hat{\rho}_{Ia}(t) = \sum_{\vq} \frac{1}{\sqrt{2\,V\,\omega_q}}\Big[  \langle b_{\vq} \rangle(t)\,e^{-i\omega_q t} +  \langle b^\dagger_{-\vq}\rangle(t)  \, e^{i\omega_q t} \Big]\,   e^{i\vq\cdot\vx}\,, \label{expvalI}  \ee where
\be  \langle b_{\vq} \rangle(t)=\mathrm{Tr}\Big(b_{\vq}\, \hat{\rho}_{Ia}(t) \Big) ~~ ; ~~\langle b^\dagger_{-\vq}\rangle(t) = \mathrm{Tr}\Big(b^\dagger_{-\vq}\, \hat{\rho}_{Ia}(t) \Big) \,. \label{aves} \ee

For any interaction picture operator $\mathcal{A}$ associated with the (ALP)  field
\be \frac{d}{dt}\langle \mathcal{A}\rangle = \mathrm{Tr}_{a}\Big\{ \dot{\mathcal{A}}~\hat{\rho}_{Ia}(t) + {\mathcal{A}}~\dot{\hat{\rho}}_{Ia}(t)\Big\} \,, \label{timederave}\ee where the average $\langle (\cdots) \rangle = \mathrm{Tr}_{a}(\cdots)\hat{\rho}_{Ia}(t)$. Because $b_{\vq} ,b^\dagger_{\vq}$ are time independent in the interaction picture, the time derivative of their expectation value is given solely by the second term on the right hand side of eqn. (\ref{timederave}), hence the expectation value of the number operator
\be N_q(t) = \mathrm{Tr}_{a}\, \hat{\rho}_{Ia}(t)\, b^\dagger_{\vq} \, b_{\vq}\,\label{numop}\ee obeys the quantum kinetic equation

\be \frac{dN_q(t)}{dt} =  \mathrm{Tr}_{a}\Big\{ b^\dagger_{\vq} \, b_{\vq} ~\dot{\hat{\rho}}_{Ia}(t)\Big\} = -\Gamma_q(t) N_q(t) + \Gamma^<_q(t) \,. \label{dNdt}\ee

Similarly, we also find the evolution equation for the averages
\bea \frac{d}{dt}\langle b_{\vk} \rangle(t)  & = &  \Big[-i\,\Delta_k(t)-\frac{\Gamma_k(t)}{2} \Big]\, \langle b_{\vk} \rangle (t)\nonumber \\\frac{d}{dt}\langle b^\dagger_{\vk}\rangle(t)  & = &  \Big[i\,\Delta_k(t)-\frac{\Gamma_k(t)}{2} \Big]\, \langle b^\dagger_{\vk} \rangle (t)\,, \label{aveaad}\eea
and for the off-diagonal coherences,
\bea \frac{d}{dt} \langle b_{\vk}~ b_{-\vk} \rangle (t) & = &  \Big[-2i\,\Delta_k(t)  - \Gamma_k(t)\Big] \langle b_{\vk}~ b_{-\vk} \rangle (t) \nonumber \\
 \frac{d}{dt} \langle b^\dagger_{\vk}~ b^\dagger_{-\vk} \rangle(t)  & = &  \Big[2i\,\Delta_k(t)  - \Gamma_k(t)\Big] \langle b^\dagger_{\vk} ~ b^\dagger_{-\vk} \rangle(t)\, .  \label{bilins}\eea From the evolution equations (\ref{aveaad},\ref{bilins}) it is clear that $\Delta_k(t)$ is a time dependent renormalization of the frequency $\omega_k$. To obtain the solutions of the above equations in the long time limit we need the following integrals

 \be \int^t_0 \Delta_q(t') dt'  =  t\,\frac{g^2}{2\,\omega_q} \int^\infty_{-\infty}   \frac{\rho(q_0,q)}{(\omega_q-q_0)} \,\Bigg[ 1-\frac{\sin(\omega_q-q_0)\,t}{(\omega_q-q_0)\,t} \Bigg]\,\frac{dq_0}{(2\pi)} ~~{}_{\overrightarrow{t\rightarrow \infty}} ~~ t\,\delta \omega_q   \label{realpartofE}  \ee where
  \be \delta \omega_q = \frac{g^2}{2\,\omega_q}\, \int^\infty_{-\infty} \mathcal{P}\Bigg[ \frac{\rho(q_0,q)}{(\omega_q-q_0)} \Bigg]\,\frac{dq_0}{2\pi}\,,\label{renfreq}\ee is a renormalization of the frequency $\omega_q$ and  $\mathcal{P}$ stands for the principal part, and
  \be  \int^t_0 \Gamma_q(t')\,dt'   =    \frac{g^2}{ \omega_q}\, \int_{-\infty}^{\infty} \frac{dq_0}{2\pi} \, \frac{\rho(q_0,q)}{( q_0-\omega_q)^2}\,\Big[ 1-\cos\big[(q_0-\omega_q)t\big] \Big]~~ {}_{\overrightarrow{t\rightarrow \infty}} ~~ \gamma_q \, t +  \frac{g^2}{ \omega_q}\,  \int_{-\infty}^{\infty} \frac{dq_0}{2\pi} \,\mathcal{P}  \frac{\rho(q_0,q)}{( \omega_q-q_0)^2}\,, \label{gamasy}\ee where
  \be \gamma_q = \Gamma_q(\infty)=\frac{g^2}{2 \omega_q}\,\rho(\omega_q,q)\,,  \label{gamafgr} \ee is the decay rate in agreement with Fermi's golden rule.
  In the long time limit, the solution of eqns. (\ref{aveaad},\ref{bilins}) are
  \be \langle b_{\vk} \rangle(t) = {Z}\,  e^{-i\delta \omega_q t}\,e^{-\frac{\gamma_q}{2}t } \,\langle b_{\vk} \rangle(0)~~;~~ \langle b^\dagger_{\vk} \rangle(t) = {Z}\,  e^{ i\delta \omega_q t}\,e^{-\frac{\gamma_q}{2}t } \,\langle b^\dagger_{\vk} \rangle(0)\,,\label{solubs}  \ee
  \be \langle b_{\vk}b_{-\vk} \rangle(t) = {Z}^2\,  e^{-2i\delta \omega_q t}\,e^{- \gamma_q t } \,\langle b_{\vk}b_{-\vk} \rangle(0)~~;~~ \langle b^\dagger_{\vk} b^\dagger_{-\vk}\rangle(t) = {Z}^2\,  e^{ 2i\delta \omega_q t}\,e^{- \gamma_q t } \,\langle b^\dagger_{\vk} b^\dagger_{-\vk} \rangle(0)\,,\label{solubilins}  \ee

  where to leading order in the coupling,
  \be Z = 1- \frac{g^2}{ 2\omega_q}\,  \int_{-\infty}^{\infty} \frac{dq_0}{2\pi} \,\mathcal{P}  \frac{\rho(q_0,q)}{( \omega_q-q_0)^2} \label{wfun} \ee is the wave function renormalization.

 If the initial averages $\langle b_{\vk} \rangle(0)=0~;~\langle b_{\vk}~ b_{-\vk} \rangle (0)=0$ such values remain as fixed points of the evolution equations. However for a ``misaligned'' initial condition (\ref{rhozeroalp},\ref{eigenb}) yielding the initial averages (\ref{trazasbb}),  it follows that   in the long time limit the solutions of eqns. (\ref{aveaad},\ref{bilins}) are, respectively
\be \langle b_{\vq} \rangle(t)  = Z\,e^{-i \delta \omega_q\,t}\, e^{-\frac{\gamma_q}{2}t }\,\Delta_{\vq}  \,, \label{soluat}\ee
\be \langle b_{\vq}~ b_{-\vq} \rangle(t)= Z^2\,e^{-2 i \delta \omega_q \,t}\, e^{-\gamma_q t}\,\Delta_{\vq}\Delta_{-\vq} \,,\label{solbilt} \ee along with their hermitian conjugates.

Absorbing   $\delta \omega_q$ into  the renormalization of the frequency and with the initial expectation values given by (\ref{expealp},\ref{expepialp}),\ref{deltazero})   we find that the expectation value of the (ALP) field is given by
\be \langle a \rangle(t) = e^{-\frac{\gamma_0}{2}\,t}\Big[\overline{a}(0)\,\cos(m_{aR}\,  t) + \frac{\overline{\pi}(0)}{m_{aR}}\,\sin(m_{aR}\, t)+ \mathcal{O}(g^2)\Big]  \,, \label{alpext}\ee where $m_{aR}$ is now the renormalized (ALP) mass and we have neglected (non-secular) terms of order $g^2$ associated with the wave function and mass renormalizations. Equations (\ref{soluat},\ref{solbilt},\ref{alpext}) indicate  that the expectation values and off-diagonal coherences   decay in time, leading to a reduced density matrix diagonal in the number representation, this is the hallmark of \emph{decoherence}. These results imply that the damping of the (ALP) condensate is directly linked to \emph{decoherence}.

Neglecting perturbatively small non-secular terms of $\mathcal{O}(g^2)$   in the long time limit yields  in this limit
\be \langle b^\dagger_{\vq}~ b_{\vq} \rangle(t)\equiv N_q(t) = N_q(0) \,e^{-\gamma_q t}+n(\omega_q)(1-e^{-\gamma_q t})  ~~;~~ N_q(0) = |\Delta_{\vq}|^2 ~~;~~  n(\omega_q)=\frac{1}{e^{\beta \,\omega_q}-1}\label{thermalasy} \ee which describes thermalization, and an exponential approach to the thermal fixed point of the quantum kinetic equation.

Taken together, the results given by (\ref{soluat},\ref{solbilt},\ref{alpext},\ref{thermalasy})  summarize some of the main results from the (QME): damping of the (ALP) condensate, decoherence and thermalization are all related,   the   decoherence rate is the same as the   thermalization rate as well as the damping rate of the misaligned component to the energy density. For $t \gg 1/\gamma_q$ the density matrix becomes diagonal in the occupation number basis and the misaligned condensate has relaxed to zero. The (ALP) has reached thermal equilibration with the bath.

  From eqns. (\ref{thermalasy},\ref{Nzero},\ref{zeromodi}) we obtain the time evolution of the (ALP) energy density neglecting a time independent zero point contribution, it  is given by

\be \mathcal{E}(t) = \frac{1}{V} \sum_{\vq} N_q(t) \omega_q = \frac{1}{2} \Big[\overline{\pi}^2+ m^2_a\,\overline{a}^2 \Big]\,e^{-\gamma_{ {0}}t} + \int \frac{d^3q}{(2\pi)^3}\,\omega_q \, n(\omega_q)\,(1-e^{-\gamma_q t}) \,. \label{alpEd} \ee   The first term in (\ref{alpEd}) describes the decay of the condensate from the misaligned initial state, whereas the second term describes the thermalization of the (ALP) degrees of freedom.

This analysis highlights that the contribution from a misaligned condensate to the energy density, thermalization with the bath and decoherence as described by the decay of the off-diagonal components in the (ALP) occupation number pointer basis,  all occur on similar time scales, which is completely determined by the relaxation rate $\gamma_q$.

The results (\ref{alpext},\ref{alpEd}) are in complete agreement with those of reference\cite{shuyang} which were obtained with a very different approach based on the non-equilibrium Schwinger-Keldysh effective action. Furthermore, the general expression for the frequency renormalization (\ref{renfreq}) and wave function renormalization (\ref{wfun}) are also in agreement with the general results found in ref.\cite{shuyang} in the strict perturbative regime, although they cannot reproduce non-perturbative aspects which are revealed by the effective action and are discussed in section (\ref{sec:discussion}).

\vspace{1mm}

\textbf{Decoherence and entropy production:} The evolution equations (\ref{soluat},\ref{solbilt}) describe the decay of the coherences, in other words,  the emergence of  \emph{decoherence}, whereby  the density matrix becomes diagonal in the pointer basis of the eigenstates of the occupation number operator $b^\dagger_q b_q$ for $t \gg 1/\gamma_q$. Furthermore, the time scale of decoherence  is similar to the relaxation rate of the misaligned component of the energy density and that of thermalization. Decoherence and the evolution towards a diagonal reduced density matrix in the occupation number basis, in turn imply entropy production. At long time when the off diagonal terms are negligible, and the reduced density matrix becomes diagonal in the occupation number basis, with thermal populations,  the total entropy becomes
 \be S(\infty) = \sum_q \Bigg\{(1+n(\omega_q))\,\ln(1+n(\omega_q))- n(\omega_q) \,\ln n(\omega_q)  \Bigg\}\,. \label{entropy} \ee     Since the initial (ALP) density matrix (\ref{rhozeroalp}) describes a pure state, hence vanishing entropy,   $S(\infty)>0 $ implies entropy production for the (ALP) as a consequence of environment-induced decoherence\cite{zurek}. This is an important bonus of the (QME) which unambiguously describes decoherence via the decay of the coherences (\ref{soluat},\ref{solbilt}), over the usual Boltzmann equation approach to thermalization, wherein entropy production is inferred via Boltzmann's H-theorem from the time  evolution of a classical $H(t)$ function which inputs solely the occupation number evolution but which does not have any information on off-diagonal coherence.

\vspace{1mm}

\section{(ALP)-photon interactions:}\label{sec:EBint}

 The results obtained in the previous section are general up to $\mathcal{O}(g^2)$ and to all orders in the couplings of the bath field $\chi$ to any other field except for the (ALP). Whereas our study addresses the non-equilibrium dynamics of (ALP) fields, the results also apply to any field with an interaction of the form  (\ref{lag}) and initial conditions that allow for the evolution of a coherent condensate\cite{turner}.   However, although the results are generic,   the relaxation rate $\gamma_q$,  frequency and wave function renormalizations  depend on the spectral properties of the bath correlations.

 In this section we focus on (ALP) interaction with photons via the coupling
 \be \mathcal{L}_I = -g a(x) \vec{E}(x)\cdot\vec{B}(x)\,. \label{alpgama}\ee

  We consider the thermal bath of (CMB) blackbody radiation of free \emph{massless} photons, neglecting electromagnetic interactions with charged leptons and quarks. This restricts the validity of our treatment to temperatures well below the masses of these other degrees of freedom and under conditions when the electron density in particular is vanishingly small, therefore there is no (gauge invariant) thermal mass  or plasma frequency for the photons. These conditions are certainly fulfilled in cosmology after recombination at temperatures $T \simeq eV$ when the free electron density vanishes rapidly and the distribution functions of quarks and charged leptons are thermally suppressed at these temperatures.

   The spectral density $\rho(q_0,\vec{q})$ has been obtained in ref.\cite{shuyang} and summarized in appendix (\ref{app:ebcoup}) for consistency of presentation, it is given by

\begin{equation}
    \rho(q_0,\vec{q})
    = \frac{(Q^2)^2}{32\pi}\,\Bigg\{\Bigg(1 + \frac{2}{\beta q}\,\ln\Bigg[\frac{1-e^{-\beta \omega^I_+}}{1-e^{-\beta \omega^I_-}} \Bigg]\Bigg)\,\Theta(Q^2)  + \frac{2}{\beta q}\, \ln\Bigg[\frac{1-e^{-\beta \omega^{II}_+}}{1-e^{-\beta \omega^{II}_-}} \Bigg]\,\Theta(-Q^2) \Bigg\}\, \mathrm{sign}(q_0)\,, \label{rhofi2}
\end{equation} where
\be Q^2= q^2_0 - q^2 ~~;~~ \omega_\pm^{(I)} = \frac{|q_0| \pm q}{2}~~;~~ {\omega}_\pm^{(II)} = \frac{q \pm |q_0|}{2}\,.\label{Q2omegas2}\ee
and $\beta = 1/T$ with $T$ the temperature of the radiation bath.

From equation (\ref{gamafgr}) we obtain the relaxation rate
\be \gamma_q(T) = \gamma_q(0) \, \Bigg(1 + \frac{2}{\beta q}\,\ln\Bigg[\frac{1-e^{-\beta \omega^I_+}}{1-e^{-\beta \omega^I_-}} \Bigg]\Bigg)_{q_0=\omega_q}~~;~~ \gamma_q(0) = \frac{g^2\,m^4_a}{64\pi\,\omega_q}\,.    \label{relarate} \ee The zero temperature contribution $\gamma_q(0)$ is recognized as the (ALP) decay rate into two photons\cite{marsh}, whereas the finite temperature contribution is a consequence of stimulated emission and absorption processes in the radiation bath. In the long-wavelength limit we find
 \be \gamma_q(T)  = \frac{g^2\,m^3_a}{64\pi} \, \Bigg(1 +  {2}\,n\Big(\frac{m_a}{2}\Big)  \Bigg)\,, \label{kzerog}\ee which in the high temperature limit $T \gg m_a$ yields
    \be  \gamma_q(T) = \frac{g^2\,m^3_a}{16\pi} \Bigg( \frac{T}{m_a}\Bigg)\,. \label{hiTg}\ee For example, if $T$ corresponds to the temperature of the cosmic microwave background today $T \simeq 10^{-4}\,\mathrm{eV}$ the finite temperature correction yields a very large enhancement over the zero temperature rate  if the (ALP) is an ultra-light candidate with $m_a \lesssim 10^{-22}\,\mathrm{eV}$. A substantial relaxation rate of the (ALP) post recombination may yield important cosmological consequences, for example in birefringence if it is caused by the coupling of (CMB) photons to a pseudoscalar (ALP)\cite{sikibiref,komatsu,mina,biref} (see discussion below).

    From  the results of appendix (\ref{app:ebcoup}),  the frequency renormalization given by eqn. (\ref{renfreq})   is found to be
    \be \delta \omega_q = \delta \omega_q^{(0)}+ \delta \omega_q^{(T)}\,, \label{delomesplit}\ee where $\delta \omega_q^{(0)}$  is obtained  from the $T=0$ contribution to the spectral density  (\ref{rhofi2}) and by introducing an ultraviolet cutoff $\Lambda$, it is given by
    \be \delta \omega_q^{(0)} = - \frac{g^2}{128\pi^2\,\omega_q}   \Big[  \frac{1}{2}  \Lambda^4  + 2m^2_a \Lambda^2  +   (m^2_a)^2\, \ln\Big[\frac{\Lambda^2\,e^{3/2}}{m^2_a}\Big]\Big]\,. \label{delomezero}\ee In appendix (\ref{app:finiTdel}) the finite temperature contribution in the high temperature limit $T\gg \omega_q$ is found to be
    \be  \delta \omega_q^{(T)}= -{\frac{g^2 \pi^2 T^4}{30\omega_q}}   \Big[ 1 +  \frac{ 15\, m^2_a}{24\,\pi^2\,T^2}+\mathcal{O}(m^4_a/T^4)+\cdots\Big]\,. \label{delomehiT} \ee The frequency renormalization  (\ref{delomesplit}) is identified as a temperature dependent \emph{mass renormalization} by writing the renormalized frequency $\Omega_q = \omega_q+ \delta \omega_q$ up to $\mathcal{O}(g^2)$ as
    \be \Omega_q= \sqrt{q^2+m^2_{R}(T)} = \sqrt{\omega^2_q+\Delta m^2(T)} = \omega_q+\frac{\Delta m^2(T)}{2\omega_q}+\cdots \equiv \omega_q + \delta \omega_q\,, \label{renofre}\ee from which we find the finite temperature renormalized mass up to $\mathcal{O}(g^2)$
    \be  m^2_R(T)= m^2_R(0)\,\Bigg[1- \frac{T^4}{T^4_c} \Bigg] ~~;~~ m^2_R(0) = m^2_a - \frac{g^2}{64\pi^2}   \Big[  \frac{1}{2}  \Lambda^4  + 2m^2_a \Lambda^2  +   (m^2_a)^2\, \ln\Big[\frac{\Lambda^2\,e^{3/2}}{m^2_a}\Big]\Big]   \,,\label{massrenoT}\ee where
    \be T_c \simeq 1.11 \sqrt{\frac{m_R(0)}{g}}\,, \label{Tc}\ee and
    we kept the leading order in the high temperature limit $T/m_a \gg 1$ in the finite temperature correction. The result (\ref{massrenoT}) agrees with ref.\cite{shuyang} which obtained a similar finite temperature mass from the non-equilibrium effective action, and indicates that $m^2_R(T)$ becomes \emph{negative} for $T> T_c$  suggesting a long wavelength instability and the possibility of an inverted phase transition as discussed in ref.\cite{shuyang}. However, within the context of the quantum master equation there is a caveat on this interpretation because the result for the frequency renormalization has been obtained in strict perturbation theory and the renormalized frequency $\omega_q+\delta \omega_q$ does not yield any instability. This caveat is discussed in more detail in section (\ref{sec:discussion}).

\vspace{1mm}

\section{Discussion and caveats}\label{sec:discussion}

\textbf{Counterrotating terms:}

In the derivation of the  quantum master equation (\ref{Linfin}) we  neglected terms of the form
\be b_{\vq}\,b_{-\vq}\,e^{-2i\omega_q t} e^{i\omega_q(t-t')} ~~;~~ b^\dagger_{\vq}\,b^\dagger_{-\vq}\,e^{2i\omega_q t} e^{-i\omega_q(t-t')}\,. \label{counter}\ee The time integral over $t'$ can be carried out following the steps leading to equation (\ref{Linfin}) yielding contributions of the form
$b_{\vq}\,b_{-\vq}\,e^{-2i\omega_q t} \rho^\lessgtr(q_0,q) \hat{\rho}_{Ia}(t)$ etc. The contribution of these terms to the equations of motion for linear or bilinear forms of $b,b^\dagger$ are straightforward to obtain, they do not yield terms that grow secularly in time because the rapid dephasing of the oscillatory terms average out in the time integrals. These are non-resonant terms and yield perturbatively small subleading contributions of the form $\delta \omega_q/\omega_q \ll 1~;~\gamma_q/\omega_q \ll 1$  in weak coupling, as compared to those obtained from equation (\ref{Linfin}) which captures the secular growth in time because of the resonances and describes the leading behavior in the long time dynamics.

\vspace{1mm}

\textbf{Factorization:}  Factorization of the full density matrix (\ref{factorization}) is one of the main assumptions in the derivation of the Lindblad form of the quantum master equation\cite{breuer,zoeller}.  This assumption  neglects correlations between (ALP) field and the thermal bath as discussed above, it may be justified for weak coupling: assuming an initial factorization, correlations will build up upon time evolution but will remain perturbatively small, hence they \emph{may} be neglected to leading order in the coupling $g$. The assumption that the total density matrix remains factorized with the bath in thermal equilibrium   which remains unaffected by the coupling to the (ALP) at all times is consistent with the interpretation of the bath as a reservoir. However, as the (ALP) population builds up as a consequence of thermalization, it is plausible that correlations between the (ALP) and the bath become stronger as the (ALP) population reaches a thermal state, leading up to  a possible breakdown of the factorization assumption. Such a scenario    merits deeper scrutiny  which is beyond the scope of this study.

\textbf{Non-equilibrium effective action vs. quantum master equation:} In reference\cite{shuyang} the time evolution of an initial density matrix was studied by implementing the in-in Schwinger-Keldysh formulation of non-equilibrium field theory. In this formulation the time evolution is described by the in-in effective action that leads to a Langevin equation of motion for the (ALP) field in terms of the retarded self energy $\Sigma$ and a noise term both related by the fluctuation dissipation relation. The solution of the Langevin equation of motion inputs the full propagator including the self-energy correction, and the (complex) poles in the propagator at
\be \omega^2_P(q) = \omega^2(q)+\Sigma(\omega_P(q);q)  \,, \label{complexpole}\ee determine the frequency and lifetime of (ALP) oscillations. The finite temperature effective mass is obtained from the real part of the solution of the equation (\ref{complexpole}) for $q=0$, namely $m^2(T)= \mathrm{Re}[\omega^2_P(q=0)]$.

 In the case of (ALP)-photon interaction with the coupling (\ref{alpgama}), after absorbing the zero temperature, ultraviolet divergent contributions into a definition of the zero temperature renormalized mass $m_{aR}$, the solution of the pole equation (\ref{complexpole}) at $q=0$ yields precisely the result (\ref{massrenoT}) to leading order in the high temperature expansion $T\gg m_{aR}$. The finite temperature mass, as properly defined by the position of the pole in the propagator at zero momentum, indicates the possibility of an instability and an inverted phase transition for $T>T_c$ as advocated in ref.\cite{shuyang}, a conclusion that does not rely on an expansion near the bare frequency.

 In contrast, the quantum master equation approach yields a \emph{perturbative} correction to the bare frequency in the form $\omega_q+\delta \omega_q$ with a real $\delta \omega_q$ which obviously does not entail any instability. The main reason for this discrepancy between the effective action and the quantum master equation can be traced to the fact that in the latter approach the time integrals in (\ref{markovlim}),  convolve the spectral representations (\ref{ggreatspec},\ref{glessspec}) with the   time dependence (\ref{ada}) featuring the external (ALP) frequency $\omega_q$. Therefore, the rates (\ref{Roftim}-\ref{gamles}) in the Lindblad (QME) (\ref{Linfin}),  are effectively evaluated at  the frequencies     $\omega_q$ yielding strictly perturbative corrections for the frequency and wave function renormalizations. At heart, this is a consequence of the perturbative nature of the (QME) in interaction picture.

Another important difference with the non-equilibrium effective action, is that as found in ref.\cite{shuyang}, the zero temperature contribution to the real part of the self energy is
\be \Sigma^{(0)}_R(\omega,q) = - \frac{g^2}{64\pi^2}  \Big[  \frac{1}{2}  \Lambda^4  + 2Q^2 \Lambda^2  +   (Q^2)^2\, \ln\Big[\frac{\Lambda^2\,e^{3/2}}{|Q^2|}\Big]\Big]~~;~~ Q^2= \omega^2-q^2\,, \label{sigcero} \ee where the logarithmic divergence multiplying $(Q^2)^2$ implies that the renormalized effective action requires a new higher derivative term $\propto (\partial^\mu \partial_\mu)^2 a^2(x)$ to absorb the logarithmic divergence from the self-energy. This is a consequence of the non-renormalizable interaction (\ref{alpgama}) since the coupling $g$ has dimensions of $(\mathrm{energy})^{-1}$.

In contrast, the (QME) yields the frequency renormalization $\delta\omega^{(0)}_q$ (\ref{delomezero}), which is proportional to the real part of the self-energy (\ref{sigcero})  evaluated on the (bare) mass shell, namely for $Q^2=m^2_a$.  Again this is a consequence of the time integrals leading to the (QME) in Lindblad form, and can be traced to the interaction picture representation of the density matrix.

Therefore,   the (QME) confirms   the damping of the misaligned (ALP) condensate, thermalization, and that the (ALP) energy density features a mixture of a ``cold'' component from the damped misaligned condensate and a ``hot'' component from thermalization, and that damping   of the cold and   thermalization  of the hot components and decoherence occur on similar time scales.

An instability and possible phase transition cannot be captured by the (QME) which relies on a perturbative expansion in interaction picture field theory, assuming a well defined
mass shell and stable oscillations of the various degrees of freedom. An instability will lead to a breakdown of most approximations: certainly the Markov and rotating wave approximations, since the former relies on a wide separation of time scales and the second on well defined mass shells associated with the oscillation frequencies.  Therefore, an instability associated with a possible phase transition and novel phases for $T>T_c$ is well beyond the realm of validity of the (QME) and should not be expected to be described reliably by it.

While the (QME)  cannot directly confirm the possibility of an inverted phase transition and the emergence of exotic phases  both  described by the Schwinger-Keldysh effective action and truly non-perturbative aspects, it does allow to understand the sources of these discrepancies in strict perturbation theory.

An important advantage of the (QME) is that it allows to obtain the time evolution of coherences and populations in a more direct manner thereby establishing that thermalization and  decoherence with  the concomitant entropy production occur on similar time scales.

\vspace{1mm}

\textbf{On the similarity of time scales:}  An important result is that the time scales of damping of the condensate (\ref{alpext}), decoherence (\ref{solbilt}) and thermalization (\ref{thermalasy}) are all very similar and simply related. This similarity originates in the form of the Lindblad (QME), eqn. (\ref{Linfin}), which solely inputs bilinears of the form $b^\dagger_q b_q$ (one annihilation and one creation operator), and this form of the (QME) unequivocally leads to the quantum kinetic equations (\ref{dNdt}-\ref{bilins}) whose solutions display the time scales of damping, decoherence and thermalization in terms of the same function $\Gamma_k(t)$. In turn, the particular form of the Lindblad (QME) is a consequence of the linear coupling of the axion to the composite operators $\mathcal{O}_\chi$ as described by the Lagrangian density (\ref{lag}). Although we have not studied non-linear axion couplings, it is quite possible that in the case of non-linear couplings the time scales could be quite different. Investigating this possibility would merit further study beyond the scope of our objectives.

\vspace{1mm}

\textbf{Possible cosmological consequences:}

Although we have studied the non-equilibrium dynamics of (ALP)'s in Minkowski space time, the results allow us to provide a preliminary extrapolation to cosmology.

A pseudoscalar (ALP)  coupled to photons as in eqn. (\ref{alpgama}) leads to cosmic birefringence, namely a frequency independent  rotation (in contrast to Faraday rotation)  in the polarization angle $\Psi$  between the surface of last scattering and today\cite{sikibiref,komatsu,mina,biref}. For a  homogeneous misaligned condensate slowly varying in time $\langle a \rangle(t)$ such a change is given by\cite{sikibiref}
\be \Delta \Psi = \frac{g}{2} \big(\langle a \rangle (t_{LSS})-\langle a \rangle(t_0)\big)\,.  \label{polapsi}\ee

The amplitude of the misaligned condensate decays as a consequence of the (ALP) interaction with the (CMB) photons, therefore the condensate is decaying during the cosmological expansion since recombination, and as described above the (ALP) fluctuations are thermalizing with the radiation bath on similar time scales. This hitherto unappreciated fact has important consequences.

If the lifetime $1/\gamma(T) \ll 1/H_0$ the amplitude of the condensate $\langle a \rangle(t_0) \simeq 0$, however, the (ALP) \emph{thermalizes} on the same time scale as the condensate decays, therefore, if the misaligned condensate has completely decayed between the surface of last scattering and today, we \emph{conjecture} that the \emph{fluctuations} of $\Psi$  would  feature a thermal power spectrum as a consequence of thermalization of axion fluctuations with the (CMB).

This \emph{conjecture} is motivated precisely by the similarity of the condensate damping and thermalization time scales revealed by the Lindblad (QME). The arguments in ref.\cite{sikibiref} leading up to the result (\ref{polapsi}) hinge on the change in the photon frequency for the different polarizations as a consequence of the coupling to the axion condensate, namely the expectation value of the axion field. However, as the solution from the Lindblad (QME) shows, the \emph{fluctuations} of the axion thermalize on the same time scale as damping of the condensate, therefore we expect that the polarization angle will feature  thermal fluctuations, since it is modified by the axion field.  Rather than focusing solely on the change in frequency for the different polarization as a consequence of the dynamical axion condensate, the dynamics of the polarization  post recombination should be described by the Stokes parameters which involve   combinations of  the transverse components  of the electric field \emph{squared}. At the quantum level the electric field is associated with a quantum operator, whose Heisenberg equation of motion involves the full axion field\cite{sikibiref,komatsu,mina,biref} both its expectation value as well as the fluctuating component. Therefore we \emph{conjecture} that the square of the electric field operator will depend on the square of the axion field
which includes the \emph{fluctuations} of the axion field. As described by the Lindblad (QME) the fluctuations  thermalize with the (CMB) on the same time scale as the mean-field (expectation value) decays. Hence, this reasoning leads us to expect that fluctuations in the Stokes parameters, which describe the polarization field, should feature a thermal spectrum. At this stage, this remains as a plausible conjecture which merits   deeper scrutiny on its own, which, however, is well beyond the original scope of this article.

 If the (ALP) lifetime is much shorter and the misaligned condensate decays prior to the last scattering
surface, then it has reached full thermalization with the  (CMB) and if it is an ultralight dark matter candidate it contributes to the effective number of ultrarelativistic degrees of freedom. If the lifetime is of the order of   $1/H_0$ then the (ALP) contributes as   ``mixed'' dark matter, with a cold component with weight $e^{-\gamma(T)/H_0} $ and a hot (thermal)  component with weight $(1-e^{-\gamma(T)/H_0} )$. This latter possibility opens a window to an interesting scenario, where the cold dark matter component would dominate at earlier time during galaxy formation and the hot component would dominate later, with a larger velocity dispersion, hence a larger free streaming length,  featuring a crossover between cold and hot components on time scales that depend on the coupling and
(ALP) mass.   This scenario brings interesting and hitherto unexplored consequences for galaxy formation that merit further study.

\section{Conclusions and further questions:}\label{sec:conclusions}

We studied the non-equilibrium dynamics of (ALP)'s motivated by the possibility that these particles belonging to a sector beyond the Standard Model may be suitable dark matter candidates. A hallmark of (ALP)'s is their coupling to pseudoscalar composite operators associated with Standard Model degrees of freedom, and in particular their coupling to electromagnetism may lead to cosmic birefringence, namely the rotation of the plane of polarization of the (CMB) with tantalizing possibilities of detection.
In this article we consider generic couplings of the (ALP) field  ($a(x)$) of the form  $g a(x) \mathcal{O}_{\chi}(x)$ where $\mathcal{O}_{\chi}(x)$ are pseudoscalar composite operators of Standard Model degrees of freedom ($\chi$)  assumed to be a bath in thermal equilibrium,  and derive a quantum master equation that describes the time evolution of the (ALP) reduced density matrix upon tracing the $\chi$ degrees of freedom. The (QME) is obtained up to $\mathcal{O}(g^2)$ but to \emph{all orders} in the coupling of the $\chi$ (Standard Model) degrees of freedom to any other degree of freedom within or beyond the Standard Model except for the (ALP). The initial (ALP) density matrix allows for a misaligned condensate. The (QME) describes the damping of the misaligned condensate, thermalization with the bath and decoherence, namely the damping of the off-diagonal reduced density matrix elements in the occupation number basis within a unified framework.

The (ALP) time dependent energy density $\mathcal{E}(t)$ features two components: a cold (c) component from the misaligned condensate and a hot (h)  component  from thermalization with the bath, with $\mathcal{E}(t) \simeq   \mathcal{E}_c \,e^{-\gamma(T)t}+\mathcal{E}_h (1-e^{-\gamma(T)t}) $ where the relaxation rate $\gamma(T)$ also describes the decoherence rate. Therefore, the damping of the misaligned condensate energy density, the approach to thermalization with the bath and decoherence all occur on the same time scales. We focus on the particular example of the (ALP) coupling to electromagnetism where $\mathcal{O}_{\chi}(x) = \vec{E}(x)\cdot\vec{B}(x)$ where the radiation field describes the (CMB) post recombination. The long wavelength relaxation rate is enhanced by emission and absorption in the photon bath and at high temperature $T \gg m_a$ and in the long wavelength limit is given by $\gamma(T)=   \frac{g^2}{16\pi}m^2_a T $ featuring a substantial enhancement over the zero temperature rate. These results are in agreement with those of ref.\cite{shuyang} but obtained with an independent method.

The time dependence of the energy density suggests that if the (ALP) is a dark matter candidate and interacts with Standard Model degrees of freedom in (local) thermal equilibrium, it provides a ``mixed'' dark matter scenario where the ``warmth'' depends on time: at earlier times it describes a cold dark matter component and at late times a hot component, with potentially profound implications on galaxy formation. If the (ALP) is an ultralight candidate, and if the misaligned condensate has completely decayed prior to the last scattering surface the thermal component  contributes to the effective number or ultrarelativistic degrees of freedom. If its lifetime is  smaller than the Hubble time, $\gamma(T) \ll 1/H_0$, the  misaligned condensate is decaying at least since after recombination and thermalizing with the (CMB) on similar time scales. Therefore, if cosmic birefringence is a consequence of the coupling of photons to a pseudoscalar (ALP), the rotation angle since the last scattering surface should include \emph{thermal} features in its fluctuation  spectrum.

These extrapolations to the cosmological setting must be taken as indicative based on the results in Minkowski space time. The next step of the program is to include cosmological expansion and assess if and how it modifies the conclusions above, we expect to report on this aspect in forthcoming studies.

\vspace{2mm}

\acknowledgements
  The authors gratefully acknowledge  support from the U.S. National Science Foundation through grant award NSF 2111743.

\appendix

\section{Environmental  correlation functions: Lehmann and spectral representations}\label{app:corre}
The dynamics and dissipative processes depend on the correlation functions $G^{\lessgtr}$   of the environment in eqn. (\ref{Linblad},\ref{ggreat},\ref{gless}).

Because the bath is in thermal equilibrium, its  initial density matrix is $\rho_\chi(0)=e^{-\beta H_{\chi}}/Tr\, e^{-\beta H_{\chi}}$ which is space-time translationally  invariant, and the Heisenberg picture operators associated with the bath are given by $\mathcal{O}_\chi(\vx,t) = e^{iH_{\chi}t}\,\mathcal{O}_\chi(\vx,0)\,e^{-iH_{\chi}t}$ we can write
\bea  G^>(\vx-\vx';t-t')  & = &  \langle \mathcal{O}_\chi(\vx,t)\mathcal{O}_\chi(\vx',t') \rangle_\chi = \int \frac{d^4k}{(2\pi)^4}~ \rho^>(\vk,k_0) e^{-ik_0(t-t')}\,e^{i\vk\cdot(\vx-\vx')} \label{Ggfd} \\
G^<(\vx-\vx';t-t')  & = &  \langle \mathcal{O}_\chi(\vx',t')\mathcal{O}_\chi(\vx,t) \rangle_\chi = \int \frac{d^4k}{(2\pi)^4}~ \rho^<(\vk,k_0) e^{-ik_0(t-t')}\,e^{i\vk\cdot(\vx-\vx')} \,. \label{Glfd} \eea These representations are obtained by  writing $\op(\vx,t) = e^{iH_\chi t}\,e^{-i\vec{P}\cdot \vx} \,\op(\vec{0},0) \,e^{-iH_\chi t}\,e^{i\vec{P}\cdot \vx}$ and introducing a complete set of simultaneous eigenstates of $H_{\chi}$ and the total momentum operator $\vec{P}$,  $(H_\chi,\vec{P})\ket{n} = (E_n,\vec{P}_n)\ket{n}$, from which we obtain the following Lehmann representations,
\begin{eqnarray}
\rho^>(k_0,\vk) & = &  \frac{(2\pi)^4}{\mathrm{Tr}e^{-\beta H_{\chi}}}~
\sum_{m,n}e^{-\beta E_n}
|\langle n| {\cal O}_\chi(\vec{0},0) |m \rangle|^2  \, \delta(k_0-(E_m-E_n))\,\delta(\vec{k}-(P_m-P_n)) \label{siggreat} \\
\rho^<(k_0,\vk) & = &  \frac{(2\pi)^4}{\mathrm{Tr}e^{-\beta H_{\chi}}}~
\sum_{m,n} e^{-\beta E_n}
 |\langle m| {\cal O}_\chi(\vec{0},0) |n \rangle|^2  \, \delta(k_0-(E_n-E_m))\,\delta(\vec{k}-(P_n-P_m))\,.
 \label{sigless}
\end{eqnarray} Upon relabelling
$m \leftrightarrow n$ in the sum in the definition (\ref{sigless}) and recalling that $\mathcal{O}_{\chi}$ is an hermitian operator,
we find the Kubo-Martin-Schwinger relation\cite{kms}

\begin{equation}
\rho^<(k_0,k)  = \rho^>(-k_0,k) = e^{-\beta k_0}
\rho^>(k_0,k)\,. \label{KMS}
\end{equation}

  The spectral density is defined as
\be \rho(k_0,k) = \rho^>(k_0,k)-\rho^<(k_0,k) = \rho^>(k_0,k)\big[ 1-e^{-\beta k_0}\big] \label{specOs}\ee
therefore
\be  \rho^>(k_0,k) = \rho(k_0,k)~\big[1+n(k_0)\big]~~;~~\rho^<(k_0,k) = \rho(k_0,k)~ n(k_0) \,, \label{relas}\ee  where
\be n(k_0) = \frac{1}{e^{\beta k_0}-1} \,. \label{bose}\ee

Furthermore, from the first equality in (\ref{KMS}) it follows that
\bea \rho(-k_0,k) & = &  - \rho(k_0,k) \,,  \label{oddros}\\ \rho(k_0,k) & > & 0 ~~\mathrm{for} ~~ k_0 > 0 \,. \label{positive}
\eea

  We emphasize that these are exact relations, the ``environmental'' fields $\chi$ may be coupled to other fields, for example, in the case of the (ALP) interaction with the electromagnetic fields,  the gauge field also interacts with electrons, charged leptons and quarks. The Lehmann representations (\ref{siggreat},\ref{sigless}) are non-perturbative and unambiguously yield the relations (\ref{KMS}-\ref{positive}) which are general, non-perturbative statements relying on thermal equilibrium and space-time translational invariance and do not depend on these couplings.

\section{Spectral density for $\vec{E}\cdot \vec{B}$ correlation functions.}\label{app:ebcoup}
We begin with the quantization  of the gauge field within a volume $V$ eventually taken to infinity,
\be \vec{A}(x)= \frac{1}{\sqrt{V}}\sum_{\vk,\lambda=1,2} \frac{\hat{\epsilon}_{\vk,\lambda}}{\sqrt{2k}}\,\Big[d_{\vk,\lambda}\,e^{-ik\cdot x}+d^\dagger_{\vk,\lambda}e^{ik\cdot x}\Big]\,, \label{aemquant}\ee  where  $\hat{\epsilon}_{\vk,\lambda}$ are the transverse polarizaton vectors chosen to be real. From  eqns (\ref{Ggfd},\ref{Glfd}) we need the correlation functions
\bea G^>(x-y) & = & \langle \vec{E}(x)\cdot \vec{B}(x)\vec{E}(y)\cdot \vec{B}(y)\rangle \,,\label{Ggeb} \\
G^<(x-y) & = & \langle \vec{E}(y)\cdot \vec{B}(y)\vec{E}(x)\cdot \vec{B}(x)\rangle = G^>(y-x) \,, \label{Gleb} \eea where  we now refer to $\langle (\cdots ) \rangle$ as averages in the thermal density matrix of free field photons.

In the thermal ensemble the expectation value $\langle \vec{E}(x)\cdot \vec{B}(x) \rangle =0$ by parity invariance.      Using Wick's theorem, the $(\vec{E}\cdot\vec{B})$ correlation function becomes
\be \langle \vec{E}(x)\cdot \vec{B}(x)\vec{E}(y)\cdot \vec{B}(y)\rangle = \sum_{i,j}\Big\{ \langle E^i(x)\, E^j(y) \rangle \langle B^i(x)\, B^j(y) \rangle + \langle E^i(x)\, B^j(y) \rangle \langle B^i(x)\, E^j(y) \rangle  \Big\}\,.  \label{correEB}\ee A straightforward calculation yields
\be  \langle E^i(x)\, E^j(y) \rangle  = \langle B^i(x)\, B^j(y) \rangle = \frac{1}{2V}\sum_{\vk} k\,\Big(\delta^{ij}-\hat{\vk}^i \hat{\vk}^j\Big)\,\Big[(1+n(k))\,e^{-ik\cdot(x-y)} + n(k) \, e^{ ik\cdot(x-y)}\Big] \,, \label{eecor}\ee similarly
\be \langle E^i(x)\, B^j(y) \rangle  = - \langle B^i(x)\, E^j(y) \rangle = -\frac{1}{2V} \sum_{\vk} k\, \Big(\hat{\epsilon}^i_{\vk,1}\,\hat{\epsilon}^j_{\vk,2}-\hat{\epsilon}^i_{\vk,2}\,\hat{\epsilon}^j_{\vk,1} \Big)\, \Big[(1+n(k))\,e^{-ik\cdot(x-y)} + n(k) \, e^{ ik\cdot(x-y)}\Big] \,,\label{ebcorr} \ee where $n(k) = 1/(e^{\beta k} -1)$. Combining the two terms in (\ref{correEB}) we find
\bea G^>(x-y) &  = &  \frac{1}{4V^2} \sum_{\vk}\sum_{\vp} kp(1-\hat{\vk}\cdot \hat{\vp})^2 \Bigg\{ \Big[(1+n(k))\,e^{-ik\cdot(x-y)} + n(k) \, e^{ ik\cdot(x-y)}\Big]\nonumber \\ & \times & \Big[(1+n(p))\,e^{-ip\cdot(x-y)} + n(p) \, e^{ ip\cdot(x-y)}\Big]\Bigg\}\,.  \label{Ggfi}\eea
Expanding the product, we perform the following change of variables in the various terms:

 \vspace{1mm}

 \textbf{1)} in the term $n(k)n(p)$: $\vk \rightarrow -\vk,\vp \rightarrow -\vp$;

  \vspace{1mm}

  \textbf{2)} in the term with $(1+n(k))n(p)$: $\vp \rightarrow -\vp$;

  \vspace{1mm}

  \textbf{3)} in the term with $n(k)(1+n(p))$: $\vk \rightarrow -\vk$.

    Taking the infinite volume limit  with $(1/V) \, \sum_{\vq} \rightarrow \int d^3q/(2\pi)^3 $ we obtain
\be G^>(x-y) = \int \frac{dq_0}{2\pi}\int \frac{d^3q}{(2\pi)^3}\,\rho^>(q_0,q)\, e^{-iq_0(t-t')}\,e^{i\vec{q}\cdot(\vx-\vy)}\,,\label{ggro} \ee where
\bea \rho^>(q_0,q) & = &  \frac{\pi}{2}\int \frac{d^3k}{(2\pi)^3}k |\vq-\vk|\Bigg\{\Big(1-\frac{\vk}{k}\cdot\frac{\vq-\vk}{|\vq-\vk|} \Big)^2\,\Big[(1+n(k))(1+n(|\vq-\vk|))  \delta(q_0-k-|\vq-\vk|)\nonumber \\ & + &  n(k)n(|\vq-\vk|)\,\delta(q_0+k+|\vq-\vk|)\Big]\nonumber \\
& + & \Big(1+\frac{\vk}{k}\cdot\frac{\vq-\vk}{|\vq-\vk|} \Big)^2\,\Big[(1+n(k))n(|\vq-\vk|)\delta(q_0-k+|\vq-\vk|) \nonumber \\ & +  & n(k)(1+n(|\vq-\vk|)) \delta(q_0+k-|\vq-\vk|)\Big]\Bigg\}\,.\label{roge} \eea Writing
\be G^<(x-y) = \int \frac{dq_0}{2\pi}\int \frac{d^3q}{(2\pi)^3}\,\rho^<(q_0,q)\, e^{-iq_0(t-t')}\,e^{i\vec{q}\cdot(\vx-\vy)}\,,\label{glro} \ee and using the relation (\ref{Gleb}) we find that $\rho^<(q_0,\vq) = \rho^>(-q_0,-\vq)$, however the sign change in $\vq$ can be compensated by $\vk \rightarrow -\vk$ inside the k-integral with the final result
\be \rho^<(q_0,\vq) = \rho^>(-q_0,\vq) \,, \label{iden}\ee  furthermore, using the identity $(1+n(w)) = e^{\beta w} n(w)$ and using the various delta functions in the definition of $\rho^>$ we find
\be \rho^<(q_0,\vq) = e^{-\beta q_0}\,\rho^>(q_0,\vq)\,, \label{inden2}\ee which is the Kubo-Martin-Schwinger relation\cite{kms}, thereby confirming the general results  (\ref{KMS}). The spectral density is given by (see eqn. (\ref{specOs})) $\rho(q_0,q) = \rho^>(q_0,q) - \rho^<(q_0,q)$ with
\bea && \rho(q_0,q)    =     \frac{\pi}{2}\int \frac{d^3k}{(2\pi)^3} \frac{1}{k w} \Bigg\{\big(k w + k^2 -\vk\cdot\vq \big)^2\,[1+n(k)+n(w)]\big(\delta(q_0-k-w)-\delta(q_0+k+w) \big) \nonumber \\ & + & \big(k w - k^2 +\vk\cdot\vq \big)^2\,(n(w)-n(k))\big(\delta(q_0-k+w)-\delta(q_0+k-w) \big)    \Bigg\}~~;~~ w = |\vq-\vk|\,.  \label{rhoeb} \eea
The spectral density is calculated by implementing the following steps:
\be \int \frac{d^3k}{8\pi^3} = \int^\infty_0 k^2 \frac{dk}{4\pi^2} d(cos(\theta)) ~~;~~ w = |\vq-\vk| = \sqrt{q^2+k^2-2kq\cos(\theta)}~~;~~ \frac{d(\cos(\theta))}{w}  = -\frac{d\,w}{kq} \,.\label{steps}\ee Carrying out the integrations, which are facilitated by the delta function constraints we find

\begin{equation}
    \rho(q_0,\vec{q})
    = \frac{(Q^2)^2}{32\pi}\,\Bigg\{\Bigg(1 + \frac{2}{\beta q}\,\ln\Bigg[\frac{1-e^{-\beta \omega^I_+}}{1-e^{-\beta \omega^I_-}} \Bigg]\Bigg)\,\Theta(Q^2)  + \frac{2}{\beta q}\, \ln\Bigg[\frac{1-e^{-\beta \omega^{II}_+}}{1-e^{-\beta \omega^{II}_-}} \Bigg]\,\Theta(-Q^2) \Bigg\}\, \mathrm{sign}(q_0)\,, \label{rhofi}
\end{equation} where
\be Q^2= q^2_0 - q^2 ~~;~~ \omega_\pm^{(I)} = \frac{|q_0| \pm q}{2}~~;~~ {\omega}_\pm^{(II)} = \frac{q \pm |q_0|}{2}\,.\label{Q2omegas}\ee

\section{Finite temperature contribution to $\delta \omega_q$}\label{app:finiTdel}

 \begin{equation}
    \delta \omega_q^{(T)}
    = \frac{g^2\,T}{64\pi^2\,q\,\omega_q}\, \mathcal{P} \int_{-\infty}^\infty   \frac{(q^2_0-q^2)^2}{\omega_q - q_0}
      \ln\Big[\frac{1 - e^{-\beta \omega_+}}{1 - e^{-\beta \omega_-}}\Big]\, dq_0
    \equiv  \frac{g^2\,T}{64\pi^2 \, q\,\omega_q}\, \mathcal{I}(q) ~~;~~  \omega_\pm = \Big|\frac{q \pm q_0}{2}\Big|\,.  \label{sigT}
\end{equation}
Since the argument of the logarithm is odd under $q_0 \rightarrow -q_0$, it follows that $\mathcal{I}$ can be written as
\be \mathcal{I}(q) =\mathcal{P} \int^\infty_0 \frac{2q_0(q^2_0-q^2)^2}{q^2_0-\omega^2_q}\,\ln\Bigg[\frac{1-e^{-\frac{\beta}{2}|q_0-q|}}{1-e^{-\frac{\beta}{2} (q_0+q)}} \Bigg]\,dq_0 \,.   \ee

Using the results
\begin{align}
    \int_0^\infty x^n \ln\Big[1 - e^{-(x+y)}\Big] dx & = - \Gamma(n+1) Li_{2+n}(e^{-y})
    \label{eqn:I1-integral-plus}
    \\
    \int_0^\infty x^n \ln\Big[1 - e^{-|x-y|}\Big] dx & = (-1)^n \Gamma(n+1) Li_{n+2}(e^{-y}) - 2 \sum_{i=0}^{[\frac{n}{2}]} \binom{n}{2i} \Gamma(1 + 2i) \zeta(2 + 2i)\, y^{n-2i}\,,
    \label{eqn:I1-integral-minus}
\end{align}
where $Li$ is the polylogarithm, along with  the identities
\be \mathcal{P} \int^{\infty}_0 \frac{dx}{x+z}\,\Bigg( - \frac{1}{n}\,e^{-n(x+y)}\Bigg) = \frac{1}{n}\,e^{-n(y-z)}\,Ei(-n z)\ee
\be \mathcal{P} \int^{k}_0 \frac{dx}{x+z}\,\Bigg( - \frac{1}{n}\,e^{-n(k-y)}\Bigg) = -\frac{1}{n}\,e^{-n(y+z)}\,\Big[ -Ei(n z)+ Ei(n(y+z)) \Big]\ee
\be \mathcal{P} \int^{\infty}_k \frac{dx}{x+z}\,\Bigg( - \frac{1}{n}\,e^{-n(x-y)}\Bigg) =  \frac{1}{n}\,e^{ n(k+z)}\, Ei(-n(y+z))\,, \ee and the   representation of the exponential integral function
\be Ei(x) = \gamma + ln(|x|) + \sum_{n=1}^{\infty} \frac{x^n}{n\,n!}\,, \ee where $\gamma$ is Euler's constant, we   find in the     high temperature limit $T \gg \omega_q$

\be  \delta \omega_q^{(T)}= -{\frac{g^2 \pi^2 T^4}{30\omega_q}}   \Big[ 1 +  \frac{ 15\, m^2_a}{24\,\pi^2\,T^2}+\mathcal{O}(m^4_a/T^4)+\cdots\Big]\,. \label{delomehiT1} \ee


\begin{thebibliography}{99}

\bibitem{PQ} R. D. Peccei and H. R. Quinn,  Phys. Rev. Lett. 38, 1440
(1977), Phys. Rev. D 16, 1791 (1977).

\bibitem{weinaxion} S. Weinberg,  Phys. Rev. Lett. 40,
223 (1978).

\bibitem{wil} F. Wilczek,  Phys. Rev. Lett. 40, 279 (1978).

\bibitem{pres}  J. Preskill, M. B. Wise, and F. Wilczek,   Phys. Lett. B 120, 127 (1983).

\bibitem{abbott}  L. F. Abbott and P. Sikivie,  Phys. Lett. B 120, 133 (1983).


\bibitem{dine}  M. Dine and W. Fischler,
Phys. Lett. B 120, 137 (1983).

\bibitem{banks} T. Banks and M. Dine,   Nuclear Physics B 479,
173 (1996).

\bibitem{ringwald} A. Ringwald, Physics of the Dark Universe, 1, 116 (2012).

 \bibitem{marsh} D.J.E. Marsh,   Phys. Rept., 643, 1  (2016);  F. Chadha-Day, J. Ellis, D. J. E. Marsh, Sci.Adv. 8, abj3618 (2022), D. J. E. Marsh, arXiv:1712.03018; A. Diez-Tejedor, D. J. E. Marsh, arXiv:1702.02116; J. E. Kim, D. J. E. Marsh, Phys. Rev. D93, 025027  (2016).

\bibitem{sikivie1} P. Sikivie, Rev. Mod. Phys. 93, 015004 (2021).

\bibitem{sikivie2} P. Sikivie, Lect. Notes in Physics 741, 19 (2008).






\bibitem{fuzzy} W. Hu, R. Barkana, and A. Gruzinov,   Phys. Rev. Lett., 85, 1158  (2000).

\bibitem{uldm} L. Hui, J. P. Ostriker, S. Tremaine, E. Witten, Phys. Rev. D 95, 043541 (2017).

\bibitem{banik} N. Banik, A. J. Christopherson, P. Sikivie, E. M. Todarello, Phys. Rev.D95, 043542 (2017). 

 \bibitem{cast} CAST collaboration, Nature Physics, 13, 584 (2017).

\bibitem{admx} ADMX Collaboration, Phys. Rev. Lett.127, 261803 (2021). 

\bibitem{graham} P. W. Graham, I. G. Irastorza, S. K. Lamoreaux, A. Lindner, K. A. van Bibber, Ann. Rev. Nucl. Part. Sci. 65, 485 (2015).

 \bibitem{turner}  M. S. Turner,   Phys. Rev. D 28, 1243 (1983);  Phys. Rev. D 33, 889 (1986).

 \bibitem{khlopov} Z. G. Berezhiani, M. Yu. Khlopov, R. R. Khomeriki, Sov.J.Nucl.Phys.52,   65 (1990); Z. G. Berezhiani, A. S. Sakharov, M. Yu. Khlopov, Sov.J.Nucl.Phys.55,  1063 (1992).

 \bibitem{sigl} P. Carenza, A. Mirizzi, G. Sigl, Phys. Rev. D101,103016 (2020).

 \bibitem{arza} A. Arza, T. Schwetz,E. Todarello,     JCAP 10 (2020) 013.

 \bibitem{dashin} D. S. Lee, K-W.Ng, Phys. Rev. D61, 085003 (2000).

 \bibitem{mottola} L. D. McLerran, E. Mottola, M. E. Shaposhnikov, Phys. Rev. D43, 2027 (1991).



 \bibitem{friction} A. Papageorgiou, P. Quílez, K. Schmitz, JHEP 01 (2023) 169.

 \bibitem{friction1} K. Choi, S. H. Im, H. J. Kim, H. Seong,  	JHEP 02 (2023) 180.

  \bibitem{buch} M. Bolz, A. Brandenburg, W. Buchmuller, Nucl.Phys.B606, 518 (2001); Erratum-ibid.B790,336 (2008).

 \bibitem{masso} E. Masso, F. Rota, G. Zsembinszki, Phys. Rev. D 66, 023004 (2002).

   \bibitem{shuyang} S. Cao, D. Boyanovsky, Phys. Rev. D 106, 123503 (2022).


 \bibitem{sikibiref} D. Harari, P. Sikivie, Phys. Lett. B 289, 67 (1992).

 \bibitem{komatsu} E. Komatsu,  Nat Rev Phys 4, 452  (2022).

 \bibitem{mina} Y. Minami, E. Komatsu, Phys. Rev. Lett.125, 221301 (2020).

\bibitem{biref}  P. Diego-Palazuelos \emph{et.al.}, JCAP 01 (2023) 044; arXiv:2203.04830.




\bibitem{breuer} N. P. Breuer, F. Petruccione, \emph{The theory of open quantum systems}, Oxford University Press, Oxford, 2007.

\bibitem{zoeller} C. Gardiner, P. Zoeller, \emph{Quantum Noise} Springer-Verlag, Berlin (2010).

\bibitem{lin} G. Lindblad, Comm. Math. Phys. 48, 119 (1976).

\bibitem{gori} V. Gorini, A. Kossakowski, E.C.D. Sudarshan, J. Math. Phys. 17, 821, (1976).

\bibitem{pearle} G. Pearle, European Journal of Physics 33, 805 (2012).

\bibitem{weinberg1} S. Weinberg, Phys. Rev. A90, 042102 (2014);  S. Weinberg, Phys. Rev. A93, 032124 (2016); S. Weinberg, Phys. Rev. A94, 042117 (2016).



 \bibitem{banks2} T. Banks, L. Susskind, M. Peskin, Nucl. Phys. B244, 125 (1984).

\bibitem{openburra} C. Burrage, C. Käding, P. Millington, J. Minář, Phys. Rev. D 100, 076003 (2019).

\bibitem{openaka}  Y. Akamatsu,     Prog.Part.Nucl.Phys. 123, 103932  (2022);  Phys. Rev. D 91, 056002 (2015).

\bibitem{openyao}  X. Yao,  	Int. J. of Mod. Phys. A, Vol. 36, No. 20, 2130010 (2021).

\bibitem{openmiura} Y, Akamatsu, T, Miura,     EPJ Web Conf. 258, 01006 (2022) (arXiv:2111.15402).

 \bibitem{openbram}  N. Brambilla, M. A. Escobedo, J. Soto, A. Vairo, Phys. Rev. D 96, 034021 (2017); N. Brambilla, M. A. Escobedo, J. Soto and A. Vairo, Phys. Rev. D97 (2018) 074009.




\bibitem{boyopen} D. Boyanovsky,  New J. Phys. 17  063017, (2015);    Phys. Rev. D 106, 045019 (2022);    Phys. Rev. D 92, 023527 (2015).

\bibitem{hollow} T. J. Hollowood, J. I. McDonald, Phys. Rev. D 95, 103521 (2017).

\bibitem{shandera}  S. Shandera, N. Agarwal, A. Kamal,  Phys. Rev. D 98, 083535 (2018).

\bibitem{berera}  S. Brahma, A. Berera, J. Calderón-Figueroa,     JHEP 08 (2022) 225



\bibitem{vennin}  T. Colas, J. Grain, V. Vennin, Eur.Phys.J C 82, 1085 (2022).

\bibitem{bartolo} A. D. Hammou, N. Bartolo, arXiv:2211.07598.


\bibitem{kms} R. Kubo, J. Phys. Soc. Jpn, \textbf{12}, 570 (1957);
 P. C. Martin and J. Schwinger, Phys. Rev.\textbf{ 115}, 1342
(1959).

\bibitem{zurek} W. H. Zurek, Rev. Mod. Phys. 75, 715 (2003);  Annalen der Physik (Leipzig) 9 (5) 855 (2000).











\end{thebibliography}
\end{document}